\documentclass[12pt]{article}
\usepackage[reqno]{amsmath}
\usepackage{bbm}
\usepackage{epsfig}
\usepackage{array}
\usepackage{float}
\usepackage{dsfont}
\usepackage{amstext}

\usepackage{a4}

\usepackage{a4wide}
\def\ra{\rightarrow}
\def\be{\begin{equation}}
\def\ee{\end{equation}}
\def\gs{\mathrel{
   \rlap{\raise 0.511ex \hbox{$>$}}{\lower 0.511ex \hbox{$\sim$}}}}
\def\ls{\mathrel{
   \rlap{\raise 0.511ex \hbox{$<$}}{\lower 0.511ex \hbox{$\sim$}}}}

\newcommand{\onbb}{neutrinoless double beta decay }
\newcommand{\ba}{\begin{array}{c}}
\newcommand{\baz}{\begin{array}{cc}}
\newcommand{\bad}{\begin{array}{ccc}}
\newcommand{\bea}{\begin{equation} \begin{array}{c}}
\newcommand{\eea}{ \end{array} \end{equation}}
\newcommand{\ea}{\end{array}}
\newcommand{\D}{\displaystyle}
\newcommand{\dms}{\mbox{$\Delta m^2_{\odot}$}}
\newcommand{\dma}{\mbox{$\Delta m^2_{\rm A}$}}
\newcommand{\meff}{\mbox{$\langle m \rangle$}}

\begin{document}

\title{\vspace{-2cm}
\hfill {\small TUM--HEP--641/06}\\
\vspace{-0.3cm} 
\hfill {\small hep--ph/0608111} 
\vskip 0.2cm
\bf 
Scaling in the Neutrino Mass Matrix 
}
\author{%\vskip -0.5cm
R.~N.~Mohapatra$^{a,b}$\thanks{email: \tt rmohapat@physics.umd.edu }~\mbox{ 
}~~and~~
W.~Rodejohann$^b$\thanks{email: \tt werner$\_$rodejohann@ph.tum.de} 
\\\\
{\normalsize \it $^a$Department of Physics and Center for String and Particle 
Theory,}\\
{\normalsize \it University of Maryland, College Park, MD-20742, USA}\\ \\ 
{\normalsize \it $^b$Physik--Department, Technische Universit\"at M\"unchen,}\\
{\normalsize \it  James--Franck--Strasse, D--85748 Garching, Germany}
}
\date{}
\maketitle
\thispagestyle{empty}
\vspace{-0.8cm}
\begin{abstract}
\noindent
In an attempt to uncover any underlying structure in the 
neutrino mass matrix, we discuss the possibility that the ratios of 
elements of its Majorana mass matrix are equal. We call this ``strong 
scaling Ansatz'' for neutrino masses and study 
its phenomenological implications. Of three possible independent 
scale invariant possibilities, only one is allowed by current data, 
predicting in a novel way  the vanishing of $U_{e3}$ and an inverted 
hierarchy with the lightest neutrino having zero mass.  
The Ansatz has the additional virtue that it is not 
affected by renormalization running.  
We also discuss 
explicit models in which the scaling Ansatz is realized.
\end{abstract}

\newpage

\section{\label{sec:intro}Introduction}
Understanding the new physics behind the neutrino observations is one of 
the main challenges for theory \cite{review} today. While it is widely 
expected that the extreme smallness of the neutrino masses is most 
likely related to the seesaw mechanism \cite{seesaw1}, i.e, 
the existence of a $B - L$ 
symmetry broken at very high scale, there is no common consensus on 
how to understand the detailed mixing pattern. Observation implies 
a particular structure in the neutrino mass matrix 
\be \label{eq:mnu}
\left(
\bad
m_{ee} & m_{e \mu} & m_{e \tau} \\
m_{e \mu} & m_{\mu\mu} & m_{\mu \tau} \\
m_{e \tau} & m_{\mu \tau} & m_{\tau \tau} 
\ea
\right) 
\equiv 
{\cal M}_\nu = U \, {\cal M}_\nu^{\rm diag}  \, U^T  ~,
\ee
whose diagonalization in the charged lepton mass basis 
gives the PMNS matrix $U$. There are several distinct approaches:
\begin{itemize}

\item[(i)] 
symmetry approach, where one postulates family symmetries 
to constrain the mass matrix elements and studies their 
consequence; 

\item[(ii)]
anarchy approach, where all 
elements of the neutrino mass matrix 
${\cal M}_\nu$ are allowed to vary at random 
and the most probable values are then confronted with experiments and 

\item[(iii)] 
top-down GUT approach, where one considers a quark-lepton 
unified grand unified framework such as $SO(10)$ and obtains predictions for
neutrino parameters.

\end{itemize}

In this paper, we consider a different approach which is located 
somewhere between the anarchy and the symmetry approaches. 
We consider the Ansatz that the
elements of the neutrino mass matrix, in the basis in which the charged 
lepton mass matrix is diagonal, obey scaling laws. According to this 
law, ratios of certain elements of ${\cal M}_{ij}$ are same. We will call 
this the ``strong scaling Ansatz'' (SSA) on neutrino masses. 
An important property of 
the SSA is that it is not affected 
by renormalization group extrapolation as we 
run it down from the seesaw scale to the weak 
scale. This property is not shared by many other Ans\"atze used in 
literature such as $\mu$--$\tau$ or other symmetries, texture 
zeros, etc. 

The first question to ask now is which matrix elements should 
one impose the scaling Ansatz to? 
If we consider scaling for a given matrix element as 
$m_{\alpha\beta}\simeq m_0 \, r^{x_\alpha+x_\beta}$, where $m_0$ is 
flavor independent, then this implies three kinds of strong scaling 
Ans\"atze of the form that 
 $\frac{m_{\alpha_i \beta}}{m_{\alpha_i \gamma}}$ is  
independent of 
the flavor $\alpha_i$. In terms of matrix elements, this means 
\be \label{eq:strongA}
({\rm A}):~~\frac{m_{e\mu}}{m_{e\tau}} = \frac{m_{\mu\mu}}{m_{\mu\tau}} =
\frac{m_{\tau \mu}}{m_{\tau\tau}} \equiv c~.
\ee
and cyclic variations of these relations, i.e., 
\be \label{eq:strongB}
({\rm B}):~~\frac{m_{ee}}{m_{e\tau}} = \frac{m_{\mu e}}{m_{\mu\tau}} =
\frac{m_{\tau e}}{m_{\tau\tau}} \equiv c'~
\ee 
and
\be \label{eq:strongC}
({\rm C}):~~\frac{m_{e e}}{m_{e\mu}} = \frac{m_{\mu e}}{m_{\mu\mu}} =
\frac{m_{\tau e}}{m_{\tau\mu}} \equiv c''~.
\ee 
These are the only three independent strong scaling Ans\"atze 
for the $3\times 3$ case. In the next Section \ref{sec:SSC}, 
we will consider the 
phenomenological implications of these strong scaling constraints, showing 
in particular that only case (A) is allowed by current data. 

Clearly, we do not know the fundamental origin of the SSA, but if one 
looks at the hierarchy of fermion masses, such flavor dependent power 
laws for masses could easily exist and we only need to make some 
assumptions about the coefficients. Alternatively, SSA 
could arise from seesaw physics, i.e., from the structure of 
the Dirac and heavy Majorana neutrino mass matrices. 
In any case, we will discuss theoretical scenarios where 
SSA can emerge in Section \ref{sec:theory}.

\section{\label{sec:SSC}Phenomenology of the Strong Scaling Ansatz}
In order to explore the implications of SSA, we start by writing down 
the usual form of the PMNS matrix:
\begin{equation} \label{eq:Upara}
 U = \left(
 \begin{array}{ccc}
 c_{12} \, c_{13} & s_{12}\, c_{13} & s_{13}\, e^{-i \delta}\\
 -c_{23}\, s_{12}-s_{23}\, s_{13}\, c_{12}\, e^{i \delta} &
 c_{23}\, c_{12}-s_{23}\, s_{13}\, s_{12}\, e^{i \delta} & s_{23}\, c_{13}\\
 s_{23}\, s_{12}-\, c_{23}\, s_{13}\, c_{12}\, e^{i \delta} &
 -s_{23}\, c_{12}-c_{23}\, s_{13}\, s_{12}\, e^{i \delta} & c_{23}\, c_{13}
 \end{array}
 \right) \, P~, 
\end{equation}
where $P = {\rm diag}(1,e^{i \alpha}, e^{i(\beta + \delta)})$ 
contains the Majorana phases. 
Let us discuss the phenomenology of SSA, starting with case (A) 
defined in Eq.~(\ref{eq:strongA}): 
\be \label{eq:mnustrong}
{\cal M}_\nu = m_0 \, 
\left( 
\bad
A & B & B/c \\[0.2cm]
B & D & D/c \\[0.2cm]
B/c & D/c & D/c^2 
\ea
\right)~.
\ee
To discuss the phenomenology of the model, we first note that the mass 
matrix in Eq.~(\ref{eq:mnustrong}) has rank 2 and 
therefore predicts one vanishing neutrino mass. 
The eigenvector of Eq.~(\ref{eq:mnustrong}) corresponding to the zero 
eigenvalue is given by
$(0,-1/\sqrt{1 + |c|^2},c/\sqrt{1 + |c|^2})^T$. Therefore, the strong scaling 
condition is only compatible with an inverted mass hierarchy, 
$U_{e3}=0$ and hence no $CP$ violation in oscillation experiments. 
Thus, SSA provides another way to understand the vanishing of 
$U_{e3}$ and one can use small deviations from 
strong scaling as a way to understand any 
possible non-zero $U_{e3}$ value, see Section \ref{sec:break} for 
more.  
Atmospheric neutrino mixing is described by $\tan^2 \theta_{23} = |1/c|^2$,  
and data requires that $|c|$ is close, but not necessarily equal, to 1. 
Let us assume for the moment that the 
parameters are all real. The non-zero masses read 
\be
m_{2,1} = \frac{m_0}{2 \, c^2} \, \left(D + c^2 \, (A + D) \pm w \right)~,
\mbox{ where } 
w = \sqrt{4 \, B^2 \, c^2 \, (1 + c^2) + \left(D + c^2 \, (D - A)\right)^2 }~.
\ee
The solar neutrino mixing angle is given by 
\be \label{eq:t12}
\sin^2 \theta_{12} = \left( 1 + \frac{4 \, B^2 \, c^2 \, (1 + c^2)}
{(A \, c^2 - (1 + c^2) \, D + w)^2} \right)^{-1}~,
\mbox{ or } 
\tan 2 \theta_{12} = \frac{2 \, B \, \sqrt{1 + c^2}}
{(c + 1/c) \, D - A}~,
\ee
and the effective mass governing \onbb is $m_0 \, |A|$. 
To have the ratio $\dms/\dma = (m_2^2 - m_1^2)/m_2^2 \simeq 1/25$, 
typically $D + c^2  \, (A + D) \ll w$, or $A \simeq -2D$, must hold. 
This is a tuning of parameters shared by many other Ans\"atze for neutrino 
mass models. 
In this case (with $c = +1$ one needs negative $A$)  
$\sin^2 \theta_{12} \simeq \frac 12 \, (1 + A/\sqrt{A^2 + 2 \, B^2})$ 
and therefore $B^2 > A^2$ in order to have $\sin^2 \theta_{12} < \frac 12$. 
This pattern of ${\cal M}_\nu$ corresponds to the approximate conservation 
of the flavor symmetry $L_e - L_\mu - L_\tau$ \cite{lelmlt}, and one 
has $\meff \simeq \sqrt{\dma} \, \cos 2 \theta_{12} = 
\sqrt{\dma} \, (1 - 2 \, \sin^2 \theta_{12})$. 
There is also a second, somewhat tuned possibility for the parameters 
$A, B, D$, namely 
$A^2,D^2 > B^2$, which corresponds 
to $\meff \simeq \sqrt{\dma}$.\\ 

One might compare SSA matrix to the case of exact $\mu$--$\tau$ 
symmetry \cite{mutau}, which also predicts $U_{e3}=0$ and 
in addition $\cos 2 \theta_{23}=0$. A mass matrix obeying 
$\mu$--$\tau$ symmetry reads 
\be \label{eq:mutau}
{\cal M}_\nu = m_0 \, 
\left( 
\bad
A & B & B \\[0.2cm]
B & D & E \\[0.2cm]
B & E & D  
\ea
\right)~.
\ee
Hence, $\mu$--$\tau$ symmetry with the additional constraint $E = D$ 
reproduces SSA when in addition $|c| = 1$ holds. 
Breaking of $\mu$--$\tau$ symmetry will in general lead to 
non-zero $U_{e3}$ and $\cos 2 \theta_{23}$, together with a correlation 
of these parameters which depends on the way the symmetry is 
broken \cite{rnm}. 
SSA in turn has in general non-zero $\cos 2 \theta_{23}$ and breaking 
of SSA (treated in Section \ref{sec:break}) will generate 
non-zero $U_{e3}$ not directly linked to the deviation 
from zero $\cos 2 \theta_{23}$. 
Thus a key test that distinguishes this model from an approximate 
$\mu$--$\tau$ symmetric model is that here we can have departure from the 
maximal atmospheric mixing angle even though $U_{e3}=0$, whereas in the 
case of generic breaking of $\mu$--$\tau$ symmetry, there is a strong 
correlation between $\theta_{23}-\frac{\pi}{4}$ and 
$U_{e3}$. 

Another comparison should be done with the flavor symmetry 
$L_e - L_\mu - L_\tau$, which is the usual symmetry to enforce an  
inverted hierarchy \cite{lelmlt}. 
The mass matrix reads 
\be \label{eq:lelmlt}
{\cal M}_\nu = m_0 \, 
\left( 
\bad
0 & A & B \\[0.2cm]
A & 0 & 0 \\[0.2cm]
B & 0 & 0  
\ea
\right)~,
\ee
and predicts vanishing 
\dms. In analogy to SSA, $m_3$ and $U_{e3}$ are zero and atmospheric neutrino 
mixing is non-maximal: $\tan^2 \theta_{23} = |A/B|^2$. However, 
solar neutrino mixing is maximal, and one requires large 
and tuned breaking of the symmetry in order to obtain a 
sufficiently non-maximal $\theta_{12}$ \cite{PR}. 
The breaking parameters have to have at least $40 \%$ the size of 
the parameters allowed by the symmetry. 
In the case of SSA one has naturally large but non-maximal solar neutrino 
mixing, as given in Eq.~(\ref{eq:t12}), and therefore does not face 
the problems of $L_e - L_\mu - L_\tau$.\\ 

It is interesting to elaborate on renormalization aspects of the 
strong scaling condition. Presumably, SSA will be a property 
introduced at the scale $M_X \ls M_{\rm GUT} \simeq 2 \cdot 10^{16}$ GeV 
of neutrino mass generation.  Its predictions have to be compared to 
the measurements performed at low scale $m_Z$. It is well-known 
that the effect of running from high to low scale can be 
taken into account by multiplying the 
$\alpha \beta$ entry of ${\cal M}_\nu$ with 
$(1 + \delta_\alpha) (1 + \delta_\beta)$, where  
$\delta_\alpha = C \, \frac{m_\alpha^2}{16 \, \pi^2 \, v^2} 
\ln \frac{M_X}{m_Z}$, with $m_\alpha$ being the charged lepton mass 
\cite{rge}. 
The parameter $C$ is given by 3/2 in the SM and by $-(1 + \tan^2 \beta)$ 
in the MSSM. Obviously, with $\delta_\tau \gg \delta_{e, \mu}$, 
this leads in the strong scaling condition (\ref{eq:strongA}) only 
to a rescaling of $c$ with $(1 + \delta_\tau)$, i.e., the three 
ratios of the mass matrix elements still have the same value, 
which is however slightly changed. The predictions 
$m_3 = 0$ and $U_{e3}=0$ are therefore invariant under the 
renormalization running. The other predictions of 
SSA are affected: for instance, if SSA at high scale imprints the ratio 
$c$ and therefore predicts $\tan^2 \theta_{23} = 1/c^2$, 
then at low scale we have 
$\tan^2 \theta_{23} = (1 - 2 \, \delta_\tau)/c^2$. 
Note that $\mu$--$\tau$ symmetry and $L_e - L_\mu - L_\tau$ are 
in general affected by renormalization running, hence one 
generically expects $U_{e3} \neq 0$ for such models.\\

\subsection{$CP$ violating case}
The fact that $U_{e3}=0$ means that there is no $CP$ violation 
in oscillation experiments. 
This can also be seen by working with invariants: 
a useful measure of $CP$ violation in neutrino oscillation is given by 
\cite{branco} 
\bea \label{eq:jcp0}
J_{CP} = \frac{1}{8} \, \sin 2 \theta_{12}\, \sin 2 \theta_{23}\, 
\sin 2 \theta_{13}\, \cos\theta_{13}\, \sin\delta = 
\frac{\D {\rm Im} \left\{ h_{12} \, h_{23} \, h_{31} \right\} }
{\D \Delta m^2_{21} \, \Delta m^2_{31} \, \Delta m^2_{32}~}~, \\[0.3cm]
\mbox{ where } h = {\cal M}_\nu\, {\cal M}_\nu^\dagger~. 
\eea
From Eq.~(\ref{eq:mnustrong}) it follows that $h_{13} = h_{12}/c^\ast$ 
and $h_{23} = c \, h_{33}$. Since $h$ is hermitian, 
one has 
\be
{\rm Im} \left\{ h_{12} \, h_{23} \, h_{31} \right\} = 
{\rm Im} \left\{  h_{12} \, h_{23} \, h_{13}^\ast \right\} = 
{\rm Im} \left\{  c^\ast \, h_{13} \, c \, h_{33} \, h_{13}^\ast \right\}
{\rm Im} \left\{  |c|^2 \, |h_{13}|^2 \, h_{33}  \right\}
= 0~.
\ee
From $h_{13} = h_{12}/c^\ast$ 
and $h_{23} = c \, h_{33}$ it also follows that 
$h_{12} /h_{13} - h_{32}/h_{33} = 0$. 
Using the parametrization of the PMNS matrix from Eq.~(\ref{eq:Upara}), 
this condition leads for an inverted mass ordering to \cite{cr}  
\be \label{eq:IH} 
|U_{e3}| = 
\frac{\frac{1}{2} 
\D \dms \, m_3^2 \, \sin 2 \theta_{12} \, \cot \theta_{23}}
{\D \dma \, (m_3^2 + \dma) + \dms \, 
(\dma - m_3^2 \, \sin^2 \theta_{12} )}~.
\ee
Since we know that $m_3=0$, it automatically follows again that 
$|U_{e3}| = 0$ (and vice versa). 

\begin{figure}[hbt]
\begin{center}
\epsfig{file=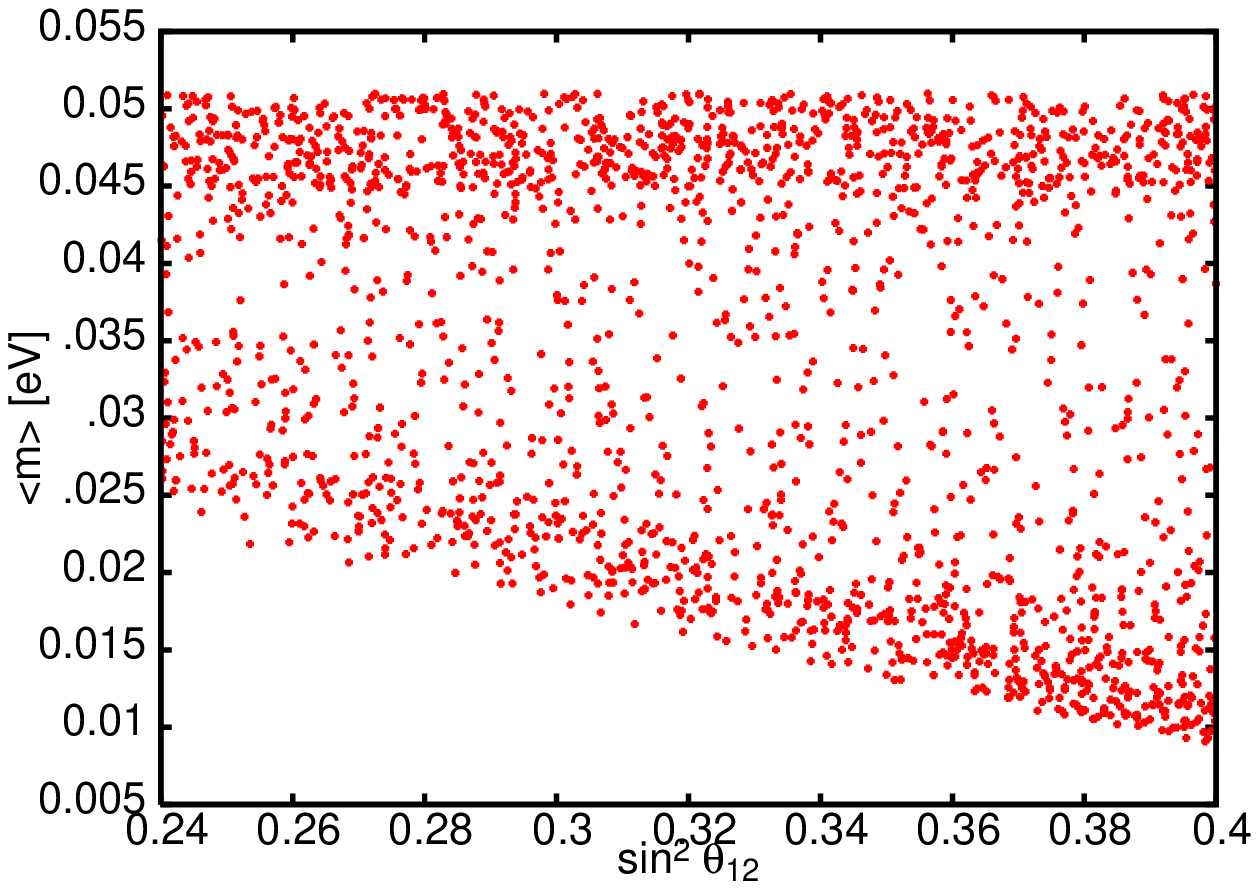,width=12cm,height=9cm}
\caption{\label{fig:strong1}Scatter plot of $\sin^2 \theta_{12}$ against 
the effective mass for Eq.~(\ref{eq:mnustrong}).}
%\end{center}
%\end{figure}
%
%\begin{figure}[hbt]
%\begin{center}
\vspace{.9cm}
\epsfig{file=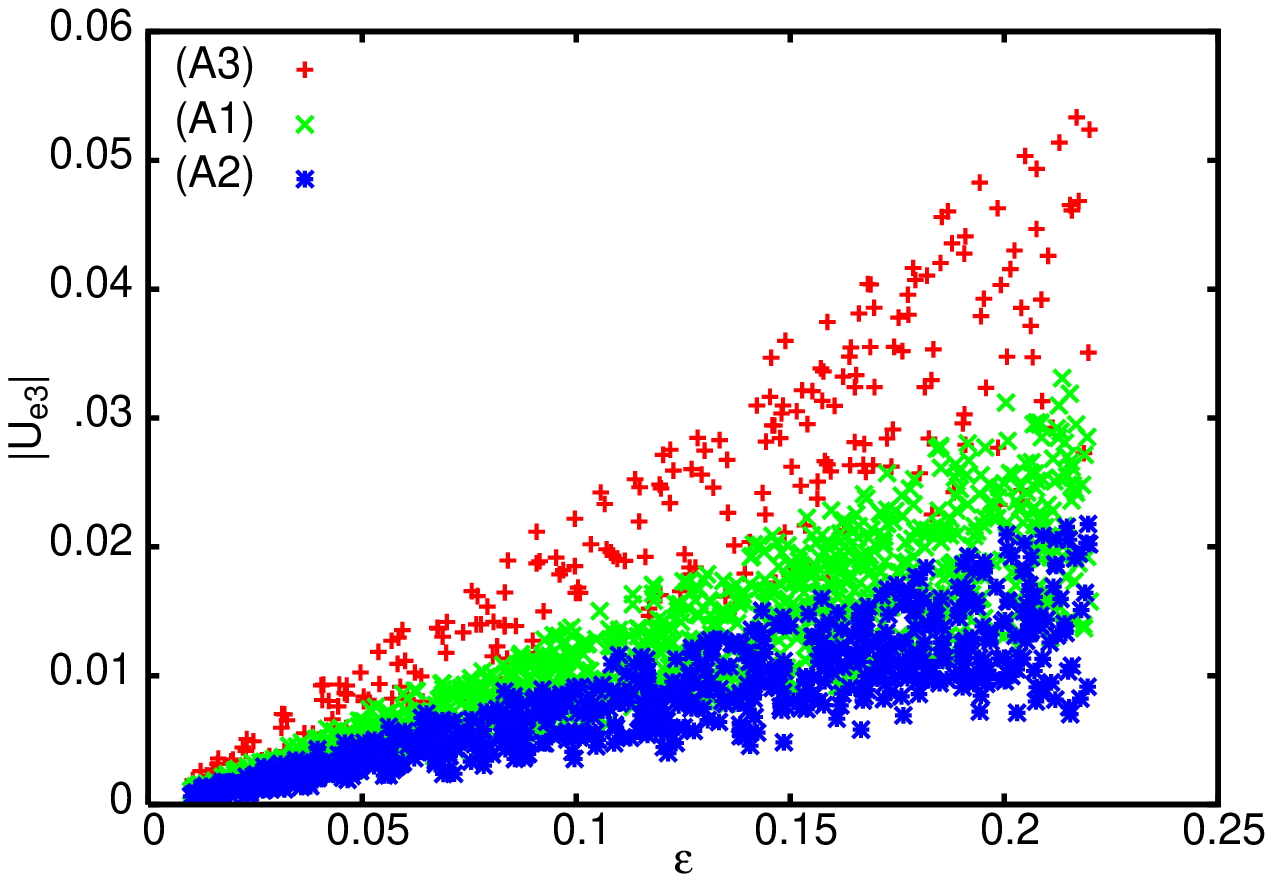,width=8cm,height=7cm}
\epsfig{file=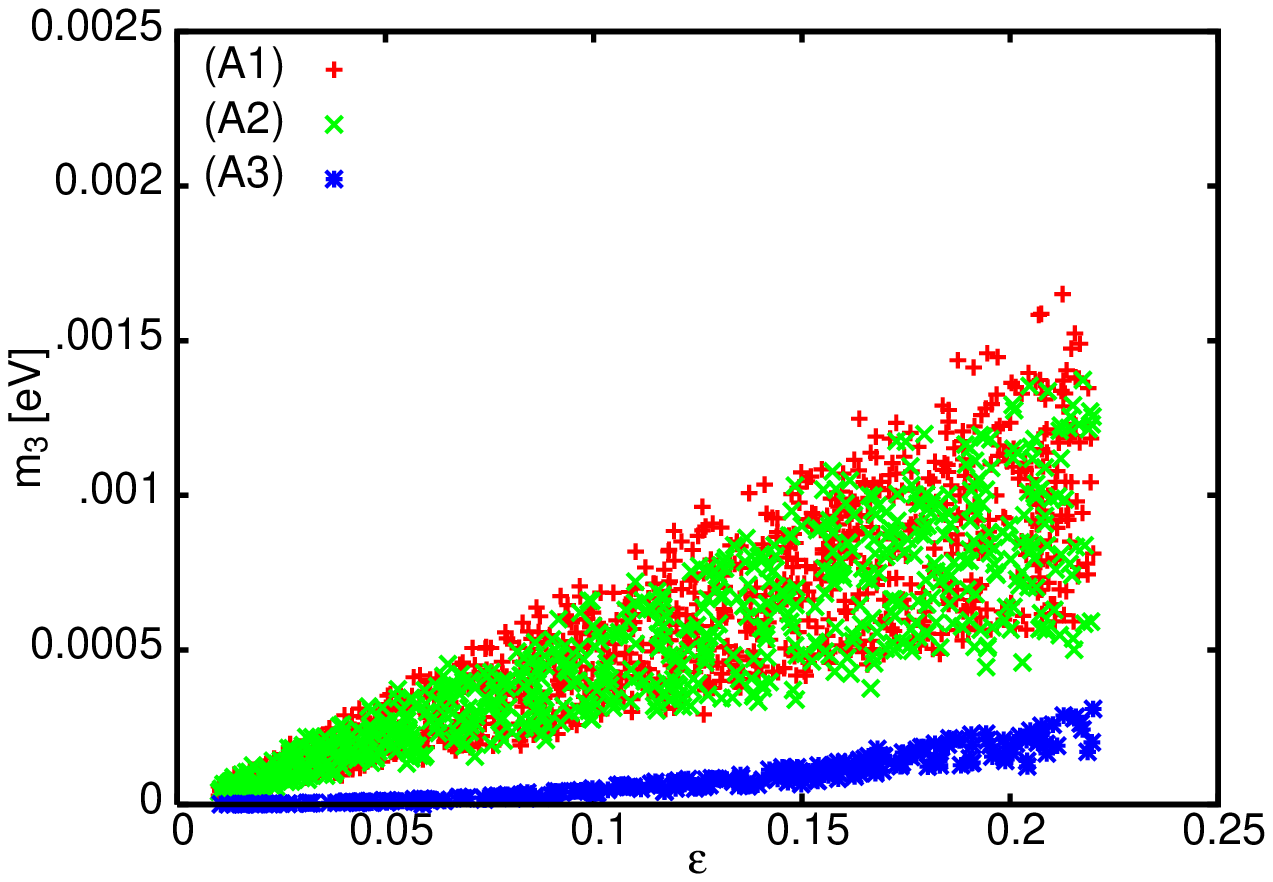,width=8cm,height=7cm}
\caption{\label{fig:break}Scatter plots of 
$|U_{e3}|$ and $m_3$ against $\epsilon$ 
for broken SSA from Eq.~(\ref{eq:break}).}
\end{center}
\end{figure}

Suppose now that all parameters in Eq.~(\ref{eq:mnustrong}) are complex, 
i.e., $A \rightarrow A \, e^{i \phi_1}$, $B \rightarrow B \, e^{i \phi_2}$, 
$D \rightarrow D \, e^{i \phi_3}$ 
and $c \rightarrow c \, e^{i \phi_4}$, where the redefined $A, B , D, c$ 
are real. It is easy to show that one can rephase the lepton fields in a way 
such that only the $e\mu$ and $e\tau$ entries are left with a phase 
$\phi = \phi_2 - \phi_1/2 - \phi_3/2$:
\be \label{eq:rephased}
{\cal M}_\nu = m_0 \, 
\left( 
\bad
A & B \, e^{i \phi} & B/c \, e^{i \phi} \\[0.2cm]
B \, e^{i \phi} & D & D/c \\[0.2cm]
B/c \, e^{i \phi} & D/c & D/c^2 
\ea
\right)~.
\ee
There are thus five relevant parameters $A, B, D, c, \phi$ for the 
observables $\theta_{12}$, $\theta_{23}$, \dms, \dma~and 
\meff. The Ansatz can therefore completely be reconstructed, in contrast 
to many other Ans\"atze.  
A 23 rotation with $\tan \theta_{23} = 1/c$ gives 
\be 
{\cal M}_\nu = m_0 \, 
\left( 
\bad
A & B \, \sqrt{1 + 1/c^2} \, e^{i \phi} & 0 \\[0.2cm]
B \, \sqrt{1 + 1/c^2} \, e^{i \phi} & D \, (1 + 1/c^2) & 0 \\[0.2cm]
0 & 0 & 0 
\ea
\right)~. 
\ee 
We note here that the SSA prediction of $m_3=0$ indicates that 
one of the Majorana phases, in this case $\beta$, is absent. 
The only observable phase is therefore the Majorana phase $\alpha$. 
With $A,B,D$ typically of the same order of magnitude, it is rather 
cumbersome to rephase this simple looking matrix in order to identify 
the sole surviving low energy phase $\alpha$ and also the two 
non-vanishing masses. Again, invariants are helpful: 
the in general three low energy phases $\alpha, \beta, \delta$ 
correspond to three independent invariants, which can be 
chosen to be \cite{utpal}
\[
I_{\alpha \beta} = {\rm Im} \left\{ m_{\alpha \alpha} 
\, m_{\beta\beta} \, 
m_{\alpha \beta}^\ast \, 
m_{\beta \alpha}^\ast \right\}~, \mbox{ where } 
(\alpha, \beta) = (e ,\mu) \mbox{ or } (e ,\tau) \mbox{ or } (\mu ,\tau)~. 
\]
In general, these expressions are rather lengthy, but for 
$m_3 = 0$ and $U_{e3}=0$ they simplify to $I_{\mu\tau} = 0$ 
and 
$I_{e \mu} = -m_1 \, m_2 \, \dms \, c_{12}^4 \, s_{12}^4 \, c_{23}^4 \, 
\sin 2 \alpha$ and 
$I_{e \tau} = -m_1 \, m_2 \, \dms \, c_{12}^4 \, s_{12}^4 \, s_{23}^4 \, 
\sin 2 \alpha$. 
The single physical low energy phase $\alpha$ corresponds 
to only one independent invariant. 
With the rephased mass matrix Eq.~(\ref{eq:rephased}) we have 
$I_{\mu\tau} = 0$, 
$I_{e \mu} = -m_0^3 \, A \, B^2 \, D \, \sin 2 \phi$ and 
$I_{e \tau} = -m_0^3 \, A \, B^2 \, D \, c^{-4} \, \sin 2 \phi$. 
By calculating the ratio $I_{e \mu}/ I_{e \tau}$ it follows again that 
$c^4 = \cot^4 \theta_{23}$. Moreover, we can write 
\bea  
m_1 \, m_2 \, \dms \, c_{12}^4 \, s_{12}^4 \, c_{23}^4 \, \sin 2 \alpha 
\simeq \dma \, \dms \, c_{12}^4 \, s_{12}^4 \, c_{23}^4 \, \sin 2 \alpha 
\\[0.3cm]
= m_0^3 \, A \, B^2 \, D \, \sin 2 \phi = \meff \, m_0^2 
\, B^2 \, D \, \sin 2 \phi~,
\eea
which shows that $\phi$ is closely related to the low energy Majorana 
phase $\alpha$. This phase $\alpha$ shows up in the effective mass 
governing \onbb$\!\!$, which in our case is just \cite{APS} 
\be \label{eq:meff}
\meff \simeq \sqrt{\dma} \, \sqrt{1 - \sin^2 2 \theta_{12} \, \sin^2 \alpha}~.
\ee
We show in Fig.~\ref{fig:strong1} a scatter plot of 
$\sin^2 \theta_{12}$ against 
the effective mass for Eq.~(\ref{eq:mnustrong}), where all parameters are 
complex. The oscillation parameters were required to lie in their current 
$3\sigma$ ranges from Ref.~\cite{thomas}. 
One can identify basically two bands in this plot, one 
corresponding to  $\meff \simeq \sqrt{\dma} \, \cos 2 \theta_{12} = 
\sqrt{\dma} \, (1 - 2 \, \sin^2 \theta_{12})$, the other to 
$\meff \simeq \sqrt{\dma}$.

Let us compare our discussion with that in the $\mu$--$\tau$ symmetric case. 
With the generic prediction of $\theta_{13}=0$, it follows 
that there is no Dirac phase. However, in contrast to the 
SSA case, in general both Majorana phases are present because there are 
in general three non-vanishing neutrino masses. 
This extra phase in $\mu$--$\tau$ case is hard to detect since 
as far as \onbb is concerned, the effective mass will depend 
only on one phase due to the fact that $U_{e3}=0$. The second phase 
is in principle observable in processes such as 
$\nu_\mu-\bar{\nu}_\mu$ and $\nu_\tau-\bar{\nu}_\tau$  
oscillations \cite{boris} or rare decays such as 
$K^+ \ra \pi^- \mu^+ \mu^+$ \cite{DL2}. The experimental challenges 
in order to observe these processes at their predicted rates 
are however depressingly breathtaking, see the discussion in Ref.~\cite{DL2}. 

For the case of $L_e - L_\mu - L_\tau$, there is also no Dirac phase, 
and in addition only one Majorana phase $\alpha$, which is fixed 
by the symmetry to the value $\pi/2$. 
The required (large and tuned) breaking of $L_e - L_\mu - L_\tau$ 
will induce non-zero $m_3$ and $\theta_{13}$ and therefore 
all three possible phases will be present. The phase $\alpha$ will 
stay close to $\pi/2$ and therefore the effective mass will stay close 
to $\meff \simeq \sqrt{\dma} \, \cos 2 \theta_{12}$, unless 
the breaking parameters are as large as 
the parameters allowed by the symmetry. In case of SSA the 
effective mass can take any value between 
$\sqrt{\dma} \, \cos 2 \theta_{12}$ and $\sqrt{\dma}$, 
which can be used to distinguish the two scenarios. We stress however 
again that $L_e - L_\mu - L_\tau$ requires unnaturally large breaking.

\subsection{Cases (B) and (C)}
Let us now discuss the cases (B) and (C) given in Eq.~(\ref{eq:strongB}) 
and Eq.~(\ref{eq:strongC}).
We note here that the effect of renormalization 
in case (B) is basically identical 
to case (A) discussed in the previous Subsection: $c$ 
gets rescaled with $(1 + \delta_\tau)$. In case (C) there 
is basically no running, because $c$ 
gets rescaled with $(1 + \delta_\mu) \ll (1 + \delta_\tau)$. 
In case (B) the mass matrix has the form:
\be
{\cal M}_\nu = m_0 \, 
\left(
\bad
A & B & A/c \\[0.2cm]
B & D & B/c \\[0.2cm]
A/c & B/c & A/c^2
\ea
\right)~.
\ee
Exactly analogous to case (A), in this case also we have a zero eigenvalue 
with an eigenvector 
given by $(-1/\sqrt{1 + |c|^2}, 0 ,c/\sqrt{1 + 
|c|^2})^T$, therefore $U_{\mu 1} = 0$ or $U_{\mu 3} = 0$, depending on 
whether the hierarchy is normal or inverted. 
This is however incompatible with observations 
which show that $U_{\mu,i} \neq 0$ for all $i= 1,2,3$. 
Similar situation holds for case (C) also, 
which predicts that  $U_{\tau 1} = 0$ or $U_{\tau 3} = 0$. 
Therefore both these cases (B) 
and (C) are not phenomenologically viable. A possible way out would be 
to take correlations from the charged lepton sector into account, which 
however requires additional input and leads to less predictivity.

\subsection{\label{sec:break}Breaking of Strong Scaling}
What happens if we break the SSA conditions? There are three possibilities 
for this, namely 
\bea \label{eq:break}\D 
({\rm A1}):~\frac{m_{e\mu}}{m_{e \tau}} = \frac{m_{\mu\mu}}{m_{\mu \tau}} = c 
~,\mbox{ but }~ \frac{m_{\tau\mu}}{m_{\tau \tau}} = c \, (1 + \epsilon) 
~,\\[0.3cm]
\D ({\rm A2}):~
\frac{m_{e\mu}}{m_{e \tau}} = \frac{m_{\tau\mu}}{m_{\tau \tau}} = c 
~,\mbox{ but }~ \frac{m_{\mu\mu}}{m_{\mu \tau}} = c \, (1 + \epsilon)
~,\\[0.3cm]
\D ({\rm A3}):~
\frac{m_{\mu\mu}}{m_{\mu \tau}} = \frac{m_{\tau\mu}}{m_{\tau \tau}} = c 
~,\mbox{ but }~\frac{m_{ee}}{m_{e \tau}} = c \, (1 + \epsilon)~.
\eea
As a consequence, non-zero $m_3$ and $U_{e3}$ will be generated and 
the prediction $\tan^2 \theta_{23} = 1/c^2$ receives corrections of 
order $\epsilon$. 
It turns out that in cases (A1) and (A2) 
$|U_{e3}|$ is roughly $\epsilon \, D/2$ and $m_3$ is roughly 
$D \, \epsilon \, \sqrt{\dma}/2$.  
In case (A3) one finds $|U_{e3}| \simeq \epsilon \, D$ and 
$m_3$ is of order $\epsilon^2 \sqrt{\dma}$. By allowing $\epsilon$ to vary 
between zero and 0.22, we display the behavior of $m_3$ and $|U_{e3}|$ 
as a function of $\epsilon$ in Fig.~\ref{fig:break}. 
The oscillation parameters were again required to lie in their current 
$3\sigma$ ranges from Ref.~\cite{thomas}. 
Only for sizable breaking of SSA of order of the Cabibbo angle one 
can probe such $|U_{e3}|$ values in next generation experiments. 
The corrections to the effective mass given in 
Eq.~(\ref{eq:meff}) are of order 
$m_3 \, |U_{e3}|^2$ and therefore completely negligible.

\subsection{\label{sec:weak}Weak Scaling?}
One could relax the strong scaling condition and introduce 
a ``weak scaling Ansatz''. 
This weaker version of SSA corresponds to the case where the equality 
holds for only a subset of flavors, e.g.,
\be \label{eq:weak1}
\frac{m_{e\mu}}{m_{e\tau}} = \frac{m_{\mu\mu}}{m_{\mu\tau}} \equiv c~,
~\mbox{ or } 
%\ee
%or
%\be \label{eq:weak2}
\frac{m_{e\mu}}{m_{e\tau}} =
\frac{m_{\tau \mu}}{m_{\tau\tau}} \equiv c~,~\mbox{ or } 
%\ee
%or
%\be \label{eq:weak3}
\frac{m_{\mu\mu}}{m_{\mu\tau}} =
\frac{m_{\tau \mu}}{m_{\tau\tau}} \equiv c~.
\ee
These relations 
are also invariant under the usual renormalization group 
extrapolations. 
Note that breaking of SSA, as discussed in the previous Subsection,  
with large $\epsilon$ corresponds to weak scaling. 
The third weak scaling condition 
has sometimes been 
invoked as a way to understand large atmospheric mixing simultaneously 
with small $\dms/\dma$ \cite{alta}.  Without additional input, for 
instance extra symmetries to make certain elements of ${\cal M}_\nu$ 
equal, there is not much predictivity for these cases. 
Therefore, we do not consider this possibility anymore.

\section{\label{sec:theory}Possible Theoretical Origin of Strong 
Scaling } 
In this Section, we speculate on the theoretical origin of the
strong scaling rule from fundamental principles. We present two gauge 
models based on discrete family symmetries, which when broken  
lead to the strong scaling rule.\\

\newpage

\noindent{\bf Scenario I: A gauge model}

First we present a gauge model, where a symmetry leads to the 
strong scaling rule. The model is based on the Standard Model gauge group 
supplemented by a family symmetry group $D_4\times Z_2$ as described in 
Ref.~\cite{cr}, where also the mathematical details of $D_4$ can be found. 
The leptons are assigned to the following representations of the discrete 
symmetry group:
\begin{eqnarray}
\{ L_e, \left(\begin{array}{c} L_\mu\\ L_\tau\end{array}\right)\}\sim 
\{1^+_1, 2^+\}~;~~\{e_R, \left(\begin{array}{c} \mu_R\\ 
\tau_R\end{array}\right)\}\sim  
\{1^-_1, 2^- \}~;~\\ \nonumber
\{(N_{e}),(N_\mu), (N_\tau)\}\sim \{1^-_1, 1^+_2,1^-_2\}~,
\end{eqnarray}
where the subscripts $+,-$ refer to the transformation under $Z_2$ and 
the rest are the $D_4$ representations. One requires \cite{cr} five Higgs 
doublets assigned to the group representations:
\begin{eqnarray}
\phi_1\sim 1^-_1~;~~ \phi_2\sim 1^+_2~;~~ \phi_3\sim 1^-_4~;~~ %\\ \nonumber
\left(\begin{array}{c}\phi_4\\ \phi_5\end{array}\right)\sim 2^+~.
\end{eqnarray}
As has been shown in \cite{cr}, this leads to the neutrino Dirac mass 
matrix of the form 
\be
m_D = 
\left( 
\bad
a & 0 & 0 \\[0.2cm]
b & d & e \\[0.2cm]
0 & 0 & 0 
\ea
\right)~, 
\ee
a diagonal charged lepton mass matrix with 
arbitrary elements and also diagonal right-handed 
neutrino mass matrices with arbitrary elements. Using these, we can 
calculate the neutrino mass 
matrix using the type I seesaw formula to get
\begin{eqnarray}
{\cal M}_\nu~=~\left(\begin{array}{ccc}
a^2\, A + b^2 \, B & b\, d \, B & b\, e \, B\\ 
b\, d \, B & d^2 \, B & d \, e \, B\\ b\, e \, B & d \, e \, B & 
e^2B\end{array}\right)~,
\end{eqnarray}
where $A,B$ are the 
inverse values of the first two right-handed neutrino masses. 
Again, strong scaling with $c = d/e$ is obeyed in this case.\\

\noindent{\bf Scenario II: 
SSA from spontaneous breaking of $\mu$--$\tau$ symmetry}

Now we present another type I seesaw model in which SSA  
follows from spontaneous breaking of discrete $\mu$--$\tau$ symmetry 
embedded into a $D_4 \times Z_2$ model. 
Consider an extension of the leptonic sector of the Standard Model with 
additional Higgs doublets $H_{e,\mu,\tau}$ and $H_{1,2}^\prime$ 
with transformation 
properties under $D_4\times Z_2$ given in Table \ref{tab:prop}. 
\begin{table}[ht]
\begin{center} 
\begin{tabular}{|c||c|} \hline 
Field & $D_4\times Z_2$ quantum number\\ \hline \hline
$L_e$, $H_e$ & $1_1^+$\\
$\left(\begin{array}{c}L_\mu\\ L_ \tau\end{array}\right)$ & $2^+$\\ 
$\left(\begin{array}{c}H_\mu\\ H_ \tau\end{array}\right)$ & $2^+$\\ 
$N_{e,\mu,\tau}$ & $1^+_1$\\
$e_R$ & $1^+_1$\\
$\left(\begin{array}{c}\mu_R\\  \tau_R\end{array}\right)$ & $2^-$\\
$H^\prime_1$ & $1^-_1$ \\
$H^\prime_2$ & $1^-_4$\\ \hline
\end{tabular}
\caption{\label{tab:prop}Transformation properties of the particle content of 
the model in Scenario IV.}
\end{center} 
\end{table}
The Yukawa couplings in this model can then be written as:
\bea
{\cal L}~=~\bar{L}_e \, H_e \, (h_1 \, N_e + h_2 \, N_\mu +h_3 \, N_\tau) 
%\\ %\nonumber 
+ (\bar{L}_\mu \, H_\mu + \bar{L}_\tau \, H_\tau) \, 
(h_4 \, N_e + h_5 \, N_\mu +h_6 \, N_\tau)\\[0.2cm] 
%\nonumber
+ f_{11} \, \bar{L}_e \, H_e \, e_R %\\ 
%\nonumber
+ f_{21} \, (\bar{L}_\mu \, H_\mu 
+ \bar{L}_\tau \, H_\tau) \, e_R +
f_{22} \, (\bar{L}_\mu \, \mu_R + \bar{L}_\tau \, \tau_R) \, 
H^\prime_1 \\[0.2cm]  %\nonumber
+ f_{33} \, (\bar{L}_\mu \, \mu_R- \bar{L}_\tau \, \tau_R) \, 
H^\prime_2~+ \frac 12 \, 
\sum\limits_{i,j} M_{ij} \, N^T_i \, C^{-1} \, N_j ~~+~h.c.
\eea
Symmetry breaking in general would lead to vevs of the form
$\langle H_{e,\mu,\tau} \rangle= v_{e,\mu,\tau}$, all being 
different from each other. 
Using the seesaw formula, we can then calculate the light neutrino mass 
matrix ${\cal M}_\nu$ and find that it satisfies SSA with 
$c=\frac{v_\mu}{v_\tau}$. Note that scaling is obeyed independent of 
the form of $M_R$. 
The atmospheric neutrino observations imply 
that $v_\mu \simeq v_\tau$ which is a constraint on the model. 

Turning to the charged lepton masses, with the definition 
$\bar{\psi}_L \, M_\ell \, \psi_R$ we can write
\begin{eqnarray}
M_\ell~=~\left(\begin{array}{ccc} f_{11} \, v_e & 0 & 0\\
f_{21} \, v_\mu & f_{22} \, v^\prime_1+f_{33} \, v^\prime_2 & 0\\ 
f_{21} \, v_\tau & 
0 & f_{22} \, v^\prime_1-f_{33} \, v^\prime_2\end{array}\right)~, 
\end{eqnarray}
where $v^\prime_{1,2}$ are the vevs of the $H^\prime_{1,2}$ fields. 
For $f_{21}=0$, this can lead to the desired diagonal charged lepton 
masses with no contribution to $U$. For non-zero $f_{21}$ one would 
expect non-zero $U_{e3}$ of the order of the 12 mixing angle in the 
matrix diagonalizing $M_\ell$. If the entries of $M_\ell$ are 
hierarchical, $(M_\ell)_{33} \simeq m_\tau \gg (M_\ell)_{22} 
\simeq m_\mu$, then $\theta_{12}^\ell \sim (M_\ell)_{21}/m_\mu$.

\section{\label{sec:sum}Summary}
We have introduced a new concept of ``strong scaling'' to the neutrino 
mass matrix in which the ratio of certain elements of the neutrino 
mass matrix 
is a constant. We find it to be a new possibility to generate zero 
$U_{e3}$ and also the inverted hierarchy. 
In contrast to the flavor symmetry $L_e - L_\mu - L_\tau$ one can 
get correct phenomenology without any breaking of the Ansatz.  
Renormalization group running from the scale of neutrino mass 
generation to low scale has basically no effect on scaling, in particular 
$U_{e3}=0$ and $m_3 = 0$ remain true for all energy scales.   
The phenomenology is predictive and testable, and the number of parameters 
equals the number of observables so that the Ansatz 
can be fully reconstructed. Observation of 
a departure of the atmospheric mixing angle from 
maximal with tiny $U_{e3}$ would distinguish this 
model from an approximate $\mu$--$\tau$ symmetric mass matrix. 
Observation of a normal hierarchy for neutrinos or 
quasi-degenerate masses will rule out our 
Ansatz.  We have also presented scenarios for the theoretical 
realization of the strong scaling Ansatz.

\vspace{0.3cm}
\begin{center}
{\bf Acknowledgments}
\end{center}
The work of R.N.M.~was supported by the National Science Foundation 
grant no.~Phy--0354401 and the 
Alexander von Humboldt Foundation (the Humboldt Research Award). 
The work of W.R.~was supported by 
the ``Deutsche Forschungsgemeinschaft'' in the 
``Sonderforschungsbereich 375 f\"ur Astroteilchenphysik'' 
and under project number RO--2516/3--1. R.N.M.\ would like to 
thank M.\ Lindner for discussions and hospitality at TUM during the time 
this work was done. We thank M.~Ratz for comments. 

{\it Note added:} 
After the paper was submitted to the arXiv, it was brought to 
our attention that there is a model in the literature \cite{grimus} 
leading to a mass matrix in our scaling Ansatz (A). It was also noted in 
Ref.~\cite{grimus} that the results $m_3=0$ and $U_{e3}=0$ are invariant 
under RGE corrections.

\clearpage 
\newpage

\pagestyle{empty}

\end{document}